\documentclass[a4paper,10pt,twoside]{cpc-hepnp}

\usepackage{multicol}
\usepackage{graphicx}
\usepackage{booktabs}
\usepackage{amssymb,bm,mathrsfs,bbm,amscd}
\usepackage[tbtags]{amsmath}
\usepackage{lastpage}

\newcommand{\bea}{\begin{eqnarray}}
\newcommand{\eea}{\end{eqnarray}}
\newcommand{\bc}{\begin{center}}
\newcommand{\ec}{\end{center}}
\newcommand{\bfi}{\begin{figure}}
\newcommand{\efi}{\end{figure}}
\newcommand{\bi}{\begin{itemize}}
\newcommand{\ei}{\end{itemize}}
\newcommand{\im}{\mbox{Im}}

\newcommand{\no}{\nonumber}
\def\ee        {\ensuremath{e^+ e^-}}

\def\ppp        {\ensuremath{p\overline{p}}}
\def\nn        {\ensuremath{N\overline{N}}}

\def\qth       {\ensuremath{q_{\rm theo}^2}}
\def\qph       {\ensuremath{q_{\rm phys}^2}}

\begin{document}



\title{Nucleon form factors:\\
the space-time connection}

\author{%
      S. Pacetti\email{simone.pacetti@pg.infn.it}%
}
\maketitle

\address{%
(INFN, Laboratori Nazionali di Frascati, Frascati, Italy)\\
(Museo Storico della Fisica e Centro Studi e Ricerche ``E. Fermi'', Rome, Italy)\\
}

\begin{abstract}
Analyticity of nucleon form factors allows to derive sum rules which, using
space-like and time-like data as input, can give unique information about
behaviors in energy regions not experimentally accessible. Taking advantage from
new time-like data on proton-antiproton differential cross section and hence the 
possibility to separate electric and magnetic form factors also in the time-like region, 
we verify the consistency of the asymptotic behavior predicted by the perturbative 
QCD for the proton magnetic form factor.
\end{abstract}

\begin{multicols}{2}
\section{Introduction}
\label{sec:intro}
Nucleon electromagnetic form factors (FFs) play a key role in
understanding the hadronic dynamics. They describe the coupling
between a photon and a nucleon pair, and represent the only 
quantities, connected to the nucleon quark structure, that can be
measured.

Physical FFs are defined as limit values for real $q^2$ ($q$
is 4-momentum transferred by the photon) of Lorentz scalar functions
that are analytic in the $q^2$ complex plane with a cut along the
real axis (time-like region), from the theoretical threshold
$\qth\equiv (2M_\pi)^2$ up to infinity. In the space-like
region ($q^2<0$), where they are real functions, FF values can
be extracted from the differential cross section of the elastic
scattering $eN\to eN$ ($N$ stands for nucleon). In the time-like
region, starting from the theoretical threshold, FFs become complex
functions, their moduli can be extracted above the physical
threshold $\qph\equiv (2M_N)^2$, studying the 
angular distribution of the $\ee\to\nn$ cross
section.

The FF asymptotic behavior predicted by the perturbative
QCD is a power law~\cite{asy} that can be obtained either in terms of
dimensional considerations, or as a consequence of a minimal
gluon exchange among the constituent quarks, needed to
share the photon transfer momentum.

More in detail, disregarding logarithmic corrections of the strong coupling
constant, the QCD power laws, in the space-like limit: $q^2\to-\infty$, 
for each nucleon FFs are
\bea
\begin{array}{l}
\left .\begin{array}{l}
F_1\sim (-q^2)^{-2}\,, \hspace{5mm} F_2(q^2)\sim (-q^2)^{-3}
\end{array}\right.\\ 
\\
\left. \begin{array}{l}
G_E(q^2)=F_1(q^2)+\displaystyle\frac{q^2}{4M_N^2}F_2(q^2)\\
G_M(q^2)=F_1(q^2)+F_2(q^2)\end{array}\right\}\sim (-q^2)^{-2}\,,\\ 
\end{array}
\label{eq:ffs}
\eea
where $F_1$ and $F_2$ are the Dirac and Pauli FFs, which account
for the non-spin flip and the spin flip part (further suppression
factor $1/q^2$) of the nucleon electromagnetic current. 
$G_E$ and $G_M$ are, instead, the so-called electric and magnetic 
Sachs FFs, in the Breit frame (for small space-like transfer momenta)
they represent the Fourier transforms of 
the charge and magnetization distributions in the nucleon.
\section{Dispersion Relations}
\label{sec:dr}
Dispersion relations (DRs) allow to connect values of an analytic function
in different regions of its domain. Taking advantage from analyticity
and vanishing asymptotic behavior of FFs (see Sec.~{\ref{sec:intro}}),
we may define the integral relation 
\bea
G(q^2)=\frac{1}{\pi}\int_{\qth}^\infty\frac{\im\,G(s)}{s-q^2}ds
\label{eq:dr-im}
\eea
with $q^2\le\qth$, for a generic FF $G(q^2)$. This DR states that
real values of $G(q^2)$ can be obtained at any $q^2<\qth$ integrating
the imaginary part over the upper edge of the time-like cut
$(\qth,\infty)$. However to use eq.~(\ref{eq:dr-im}), in case
of nucleons, we have to face two serious issues:
\bi
\item the imaginary part of FFs is not measurable, one
      should relay in phenomenological and non-rigorous techniques to 
      extract it from cross section data;
\item even though the imaginary part was obtained from the data, 
      its values could cover only a portion of the integration interval,
      starting from the physical threshold \qph. The so-called 
      ``unphysical region'', $(\qth,\qph)$, where we expect the main
      contributions from intermediate light-meson resonances,
      is not experimentally accessible.
\ei
To get around these problems we use an idea proposed for the first
time in 1974 in Ref.~\cite{ioffe}.
\section{Sum rule}
\label{sec:sumrule}
The idea~\cite{ioffe} consists in using the DR of 
eq.~(\ref{eq:dr-im}) for the function:
\bea
\phi(z)=A_L(z,s)\frac{\ln G(z)}{\sqrt{\qth-z}},
\label{eq:phi}
\eea
where $A_L(z,s)$, with $s$ real and positive, is an 
analytic function used to suppress the FF contribution 
in the time-like region $(0,s)$, and it can be chosen 
requiring
\bea
\int_0^{s}A_L(z,s)^2dz\ll 1\,.
\no
\eea 
Following the suggestion given in Ref.~\cite{ioffe} we
use the definition 
\bea
A_L(z,s)=\sum_{l=0}^L\frac{2l+1}{(L+1)^2}P_l\left(1-
2\frac{\sqrt{s}-\sqrt{z}}{\sqrt{s}+\sqrt{z}}\right)\,,
\label{eq:al}
\eea
where $P_l$ is the Legendre polynomial of degree $l$,
while the upper limit $L$ represents an 
``attenuation-power indicator''. Following the definition 
of eq.~(\ref{eq:al}), $A_L(z,s)$ is analytic in $z$
with a cut along the whole negative real axis. 
The imaginary part of $\phi(x)$ is then ($x$ is real)
\bea
\im\,\phi(x)=\left\{
\begin{array}{ll}
\displaystyle\frac{\im\,A_L(x,s)\ln G(x)}{\sqrt{\qth-x}}
& x\le 0\\
&\\
0 & 0<x<\qth\\
&\\
\displaystyle\frac{A_L(x,s)\ln |G(x)|}{\sqrt{x-\qth}}
& x\ge \qth\,.\\
\end{array}\right.
\label{eq:im-part}
\eea
We consider the proton magnetic FF normalized to its magnetic moment 
$\mu_p$, i.e.: $G(q^2)=G_M^p(q^2)/\mu_p$, having
no poles (from analyticity), nor zeros (our assumption)
in the $q^2$ complex plane, the $\phi(z)$ function 
(see eq.~(\ref{eq:phi})) is still analytic and the DR of 
eq.~(\ref{eq:dr-im}) now reads
\bea
\phi(q^2)=\frac{1}{\pi}\int^{0}_{-\infty}\frac{\im\,\phi(t)}{t-q^2}dt
+\frac{1}{\pi}\int_{\qth}^\infty\frac{\im\,\phi(s)}{s-q^2}ds\,,
\no
\eea
for $0<q^2<\qth$.
In particular at $q^2=0$, having the normalizations $G(0)=1$ and
$\phi(0)=0$, the previous relation becomes the identity
\bea
\int^{0}_{-\infty}\frac{\im\,\phi(t)}{t}dt=
-\int_{\qth}^\infty\frac{\im\,\phi(s)}{s}ds\,,
\eea
in terms of $A_L$ and FF, i.e. using
the definition of eq.~(\ref{eq:im-part}), we have
\bea
\int_{-\infty}^0\!\!\!\!\!\!\frac{\im A_L(t,s)\ln G(t)}{t\sqrt{\qth-t}}dt
&\!\!\!\!\!\!=\!\!\!\!\!\!&
\!\!\!-\!\!\!\int_{\qth}^{\infty}\!\!\!\!\!\!\frac{A_L(s',s)\ln|G(s')|}{s'\sqrt{s'-\qth}}ds'
\no\\
&&\label{eq:eq}\\
&\!\!\!\!\!\!\simeq\!\!\!\!\!\!& 
\!\!\!-\!\!\!\int_{s}^{\infty}\!\!\!\!
\frac{A_L(s',s)\ln|G(s')|}{s'\sqrt{s'-\qth}}ds'\,.
\no
\eea
The approximate identity of eq.~(\ref{eq:eq}) holds thanks to the attenuation
in the region $(0,s)$ provided by the function $A_L(z,s)$.

Finally the sum rule we want to use is obtained from eq.~(\ref{eq:eq})
in the special case with $s=\qph$
\bea
\int_{-\infty}^0\!\!\!\!\!\!\frac{\im A_L(t,\qph)\ln G(t)}{t\sqrt{\qth-t}}dt
\simeq&&\no\\
&&\hspace{-25mm}\!\!\!-\!\!\!\int_{\qph}^{\infty}\!\!\!\!\!\!
\frac{A_L(s',\qph)\ln|G(s')|}{s'\sqrt{s'-\qth}}ds'\,.
\label{eq:sumrule}
\eea
This identity involves only measurable quantities, i.e.: real values
of the proton magnetic FF in the space-like region (left-hand side),
and modulus, only from the physical 
threshold, in the time-like region (right-hand side).
\section{Check for the asymptotic behavior}
\label{sec:check}
We verify the compatibility of space and time-like data
on $G_M^p(q^2)$ with the QCD power law behavior given
in eq.~(\ref{eq:ffs}), using the ``space-time connection''
provided by the sum rule of eq.~(\ref{eq:sumrule}).

More in detail, in the space-like region we define
the real FF as a combination of a fit of several data 
sets~\cite{sl-data} (247 data points) and a power law, i.e.
\bea
G_{\rm SL}(q^2)=\left\{
\begin{array}{ll}
G^{\rm fit}_{\rm SL}(q^2) & q^2_{\rm min} \le q^2\le 0\\
&\\
G^{\rm fit}_{\rm SL}(q^2_{\rm min})(q^2_{\rm min}/q^2)^n
& q^2 \le q^2_{\rm min}\,,\\
\end{array}\right.\label{eq:sl}
\eea
where $q_{\rm min}^2\sim -30\,{\rm GeV}^2$ is the energy of the lower data point.
 
In the time-like region the situation is more troublesome,
indeed 
all data are on the modulus of an effective FF
which corresponds to $|G_M^p(q^2)|$ only when $|G_M^p(q^2)|=|G_E^p(q^2)|$,  
but that happens, by definition (see eq.~(\ref{eq:ffs})), solely at $q^2=\qph$!
Hence, to extract genuine values of $|G_M^p(q^2)|$ we used
the BaBar data on the $\ee\to\ppp$ total cross section and angular 
distribution~\cite{tl-data} together with
a dispersive technique to disentangle $|G_E^p(q^2)|$ and
$|G_M^p(q^2)|$~\cite{noi}. 

Similarly to what we did in the space-like region,
the modulus of the FF in the time-like region is defined as
\bea
G_{\rm TL}(q^2)=\left\{
\begin{array}{ll}
G^{\rm fit}_{\rm TL}(q^2) & 0 \le q^2\le q^2_{\rm max}\\
&\\
G^{\rm fit}_{\rm TL}(q^2_{\rm max})(q^2_{\rm max}/q^2)^n
& q^2 \ge q^2_{\rm max}\,,\\
\end{array}\right.
\label{eq:tl}
\eea
where $q_{\rm max}^2\sim 20\,{\rm GeV}^2$ is the energy of the higher BaBar data point.

In both definitions of eq.~(\ref{eq:sl}) and eq.~(\ref{eq:tl}) we used the
same power law as a consequence of the Phragm\`en-Lindel\"of theorem~\cite{pl}.
Such a theorem states that, not only the power $n$ which rules the asymptotic 
behavior, but also the limit must be the same, i.e.
\bea
\lim_{q^2\to -\infty}G_{\rm SL}(q^2)=\lim_{q^2\to +\infty}G_{\rm TL}(q^2)\,.
\no
\eea 

It follows that the sum rule of eq.~(\ref{eq:sumrule}), once all the
theoretical and experimental information have been considered, becomes an 
equation with only one unknown the power $n$, and the result we obtained
is
\bea
n=2.27\pm 0.36\,.
\no
\eea
Figure~\ref{fig:ris} summarizes this result. The shaded bands represent:
the fit functions in the data regions, and the power laws at high space
and time-like energies ($q^2< -30\,{\rm GeV}^2$ and $q^2> 20\,{\rm GeV}^2$). 
The lined central area is the unphysical region, which does not contribute
to the sum rule, being suppressed by the function $A_L(\qph,z)$.
\begin{center}
\includegraphics[width=80mm]{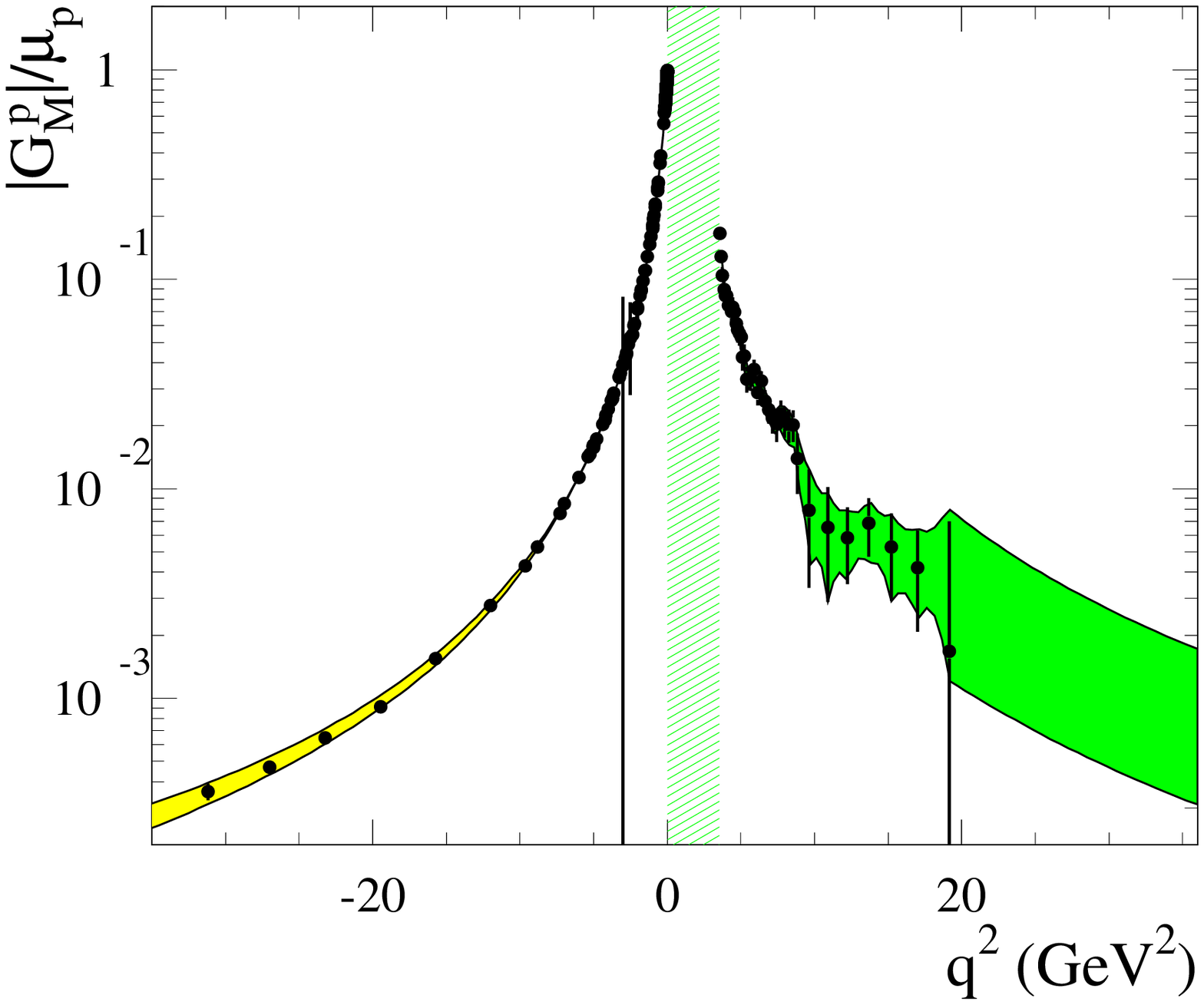}
\figcaption{\label{fig:ris} Modulus of the normalized proton magnetic FF in space-like
(left) and time-like (right) region. The bands represent the obtained 
description (see text), the points are the data~\cite{sl-data,tl-data,noi}. 
The lined central area is the unphysical region.}
\end{center}
In conclusion, using the sum rule of eq.~(\ref{eq:sumrule}), based on 
analyticity properties of FFs, we have shown that experimental data
in space and time-like region are consistent with the QCD asymptotic
behavior. In particular, we found a power law for $G_M^p(q^2)$ which is 
in good agreement with the perturbative QCD expectation.   
%
%
%
%
%
%
\end{multicols}
\vspace{-2mm}
\centerline{\rule{80mm}{0.1pt}}
\vspace{2mm}
\begin{multicols}{2}

\end{multicols}

\clearpage

\end{document}